\documentstyle[11pt]{article}

\def\kon#1#2{\vbox{\halign{##&&##\cr\lower4pt
\hbox{$\scriptscriptstyle\vert$}\hrulefill &\hrulefill\lower4pt
\hbox{$\scriptscriptstyle\vert$}\cr $#1$&$#2$\cr}}}

\def\fii{\varphi}
\def\al{\alpha}

\def\eh{{\scriptstyle{1\over 2}}}
\def\d{\partial}
\def\=d{\,{\buildrel\rm def\over =}\,}

\def\sqr#1#2{{\vcenter{\vbox{\hrule height.#2pt\hbox{\vrule width.
#2pt height#1pt \kern#1pt \vrule width.#2pt}\hrule height.#2pt}}}}
 
\def\eps{\varepsilon}

\def\te{\vartheta}
\def\fii{\varphi}
\def\pe{{\cal P}}
\def\al{\alpha}

\begin{document}

\title{Schwarzschild geodesics in terms of elliptic functions and the related red shift}
\author{\\G\"unter Scharf
\footnote{e-mail: scharf@physik.unizh.ch}
\\ Institut f\"ur Theoretische Physik, 
\\ Universit\"at Z\"urich, 
\\ Winterthurerstr. 190 , CH-8057 Z\"urich, Switzerland}

\date{}

\maketitle\vskip 3cm

\begin{abstract} Using Weierstrassian elliptic functions the exact geodesics
in the Schwarzschild metric are expressed in a simple and most transparent
form. The results are useful for analytical and numerical applications.
For example we calculate the perihelion precession and the light deflection
in the post-Einsteinian approximation. The bounded orbits are
computed in the post-Newtonian order. As a topical application we calculate
the gravitational red shift for a star moving in the Schwarzschild field.

\end{abstract}

\newpage

\section{Introduction}

Schwarzschild geodesics are elliptic functions, therefore, they should be written
as such. For this purpose the Weierstrassian elliptic functions are most useful because
they lead to simple expressions. The reason for this is that the solution of quartic
or cubic equations can be avoided in this way.

In a recent paper [1] an analytic solution for the geodesic in the weak-field
approximation was given. As pointed out in that paper the progress in the astronomical
observations call for better analytical methods. In this respect it is desirable to
have the exact geodesics in a form most suited for applications. For the orbits in
polar coordinates (next section) this goal can be achieved by using Weierstrass' 
$\pe$-function for which many analytical and numerical methods are known [2].
Considering the motion in time (section 4) the related $\zeta$- and $\sigma$-functions
of Weierstrass appear.

Jacobian elliptic functions have been used by Darwin [8] for the form of the orbits.
After some transformation our result (2.13) agrees with his. But in his second paper
he abandons the elliptic functions because they were ``not so well adapted to a study
of the time in those orbits''. Obviously the Weierstrass functions are better suited
for the problem. Indeed, expressing them by Theta functions one gets the natural expansion
of the geodesics in powers of the Schwarzschild radius, this expansion involves
elementary functions only. The Weierstrass functions have also been used by Hagihara [11]
\footnote{ I am indebted to C. L\"ammerzahl and P. Fiziev for bringing this reference
to my attention}. But he has chosen the variables and constants of integration
in a manner which leads to less explicit results. So it is difficult to derive
the post-Newtonian corrections to the geodesics given here from his formulas.
 As a topical application we finally calculate the red shift 
for a star moving in the Schwarzschild field. The geodesics are also needed for the
study of modifications of general relativity ([10], section 5.13).

\section{The orbits in polar coordinates $r=r(\fii)$}

We take the coordinates
$x^0=ct$, $x^1=r$, $x^2=\te$, $x^3=\fii$ and
write the Schwarzschild metric in the form
$$ds^2={r-r_s\over r} c^2dt^2-{r\over r-r_s}dr^2-r^2(d\te^2+\sin^2\te d\fii^2)\eqno(2.1)$$
where $r_s={2GM\over c^2}$ is the Schwarzschild radius. We shall assume $c=1$ in the
following. The geodesic equation
$${d^2x^\al\over ds^2}+\Gamma^\al_{\beta\gamma}{dx^\beta\over ds}{dx^\gamma\over ds}
=0\eqno(2.2)$$
with the Christoffel $\Gamma^\al_{\beta\gamma}$ leads to the following three
differential equations
$${d^2t\over ds^2}+\nu'{dt\over ds}{dr\over ds}=0\eqno(2.3)$$
$${d^2r\over ds^2}+{\nu'\over 2}e^{2\nu}\Bigl({dt\over ds}\Bigl)^2-
{\nu'\over 2}\Bigl({dr\over ds}\Bigl)^2-re^{\nu}\Bigl({d\fii\over ds}
\Bigl)^2=0\eqno(2.4)$$
$${d^2\fii\over ds^2}+{2\over r}{dr\over ds}{d\fii\over ds}=0.\eqno(2.5)$$
Here we have used the standard representation
$$1-{r_s\over r}=e^\nu\eqno(2.6)$$
and have chosen $\te=\pi/2$ as the plain of motion. The Christoffel symbols can
be taken from the Appendix of [3].

Multiplying (2.3) by $\exp\nu$ we find
$${\d\over ds}\Bigl(e^{\nu}{dt\over ds}\Bigl)=0$$
so that
$$e^{\nu}{d\,t\over ds}={\rm const.}= E$$
$${dt\over ds}=Ee^{-\nu}.\eqno(2.7)$$
Next multiplying (2.5) by $r^2$ we get
$$r^2{d\fii\over ds}={\rm const.}=L,$$
hence
$${d\fii\over ds}={L\over r^2}.\eqno(2.8)$$
For the constants of integration we use the notation of Chandrasekhar [4].

Finally, substituting (2.7) and (2.8) into (2.5) and multiplying by
$2\exp(-\nu)$ $\times dr/ds$ we obtain
$${d\over ds}\Bigl[e^{-\nu}\Bigl({dr\over ds}\Bigl)^2-E^2e^{-\nu}+
{L^2\over r^2}\Bigl]=0.\eqno(2.9)$$
Consequently, the square bracket is equal to another constant $=b$.
Then the resulting differential equation can be written as
$$\Bigl({dr\over ds}\Bigl)^2=E^2+e^{\nu}\Bigl(b-
{L^2\over r^2}\Bigl).\eqno(2.10)$$
The constant $b$ can be arbitrarily adjusted by rescaling the affine parameter $s$.
Below we shall take $b=-m^2$ where $m$ is the rest mass of the test particle.
This will enable us to include null geodesics (light rays) with $m^2=0$.
Each geodesic is characterized by two constants of the motion: energy $E$ and
angular momentum $L$.

Taking the square root of (2.10) and dividing by (2.8) we get
$${dr\over d\fii}=\sqrt{{E^2-m^2\over L^2}r^4+{m^2\over L^2}r_sr^3-r^2
+r_sr}\equiv \sqrt{f(r)}.\eqno(2.11)$$
Now $\fii=\fii(r)$ can be written as an elliptic integral. However, it is better
to consider the inverse $r=r(\fii)$ in terms of elliptic function by using a
formula of Weierstrass ([5], p.452). Let the quartic $f(r)$ be written as
$$f(r)=a_0r^4+4a_1r^3+6a_2r^2+4a_3r+a_4,\eqno(2.12)$$
and let $r_1$ be a zero $f(r_1)=0$, then a solution of (2.11) is given by
$$r=r_1+{f'(r_1)\over 4\pe(\fii;g_2,g_3)-f''(r_1)/6}.\eqno(2.13)$$
Here $\pe(\fii;g_2,g_3)$ is Weierstrass' $\pe$-function with invariants
$$g_2=a_0a_4-4a_1a_3+3a_2^2\eqno(2.14)$$
$$g_3=a_0a_2a_4+2a_1a_2a_3-a_2^3-a_0a_3^2-a_1^2a_4.\eqno(2.15)$$
In our case we have $a_4=0$. For the convenience of the reader we reproduce the
short proof in the Appendix.

The result (2.13) is not yet the solution of our problem because it contains too
many constants: the invariants $g_2, g_3$ and the derivatives of $f$ depend on $E, L$,
but in addition the zero $r_1$ appears. Of course one could calculate $r_1$ as a
function of $E, L$ by solving the quartic equation $f(r)=0$, but this gives
complicated expressions. It is much better to use $r_1$ and a second zero $r_2$
as constants of integration instead of $E, L$. This is even desirable from the
astronomers point of view because the zeros of derivative (2.11) are turning points
of the geodesic, for example in case of a bounded orbit they can be identified with
the perihelion and aphelion of the orbit. In order to express $E, L$ by $r_1, r_2$
we write our quartic in the form
$$f(r)=a_0r(r-r_1)(r-r_2)(r-r_3)\eqno(2.16)$$
and compare the coefficients of $r^3, r^2, r$ with (2.12). This leads to
$$4a_1=-a_0(r_1+r_2+r_3)={m^2\over L^2}r_s$$
$$6a_2=a_0(r_1r_2+r_1r_3+r_2r_3)=-1\eqno(2.17)$$
$$4a_3=-a_0r_1r_2r_3=r_s.$$
Since
$$a_0={E^2-m^2\over L^2}\eqno(2.18)$$
we can solve for
$${m^2\over L^2}r_s={r_1+r_2+r_3\over r_1r_2+r_1r_3+r_2r_3},\eqno(2.19)$$
$$E^2-m^2=-{m^2r_s\over r_1+r_2+r_3}.\eqno(2.20)$$
In addition we obtain the third zero
$$r_3=r_s{r_1r_2\over r_1r_2-r_1r_s-r_2r_s}.\eqno(2.21)$$

The relations (2.19-21) allow to express everything in terms of $r_1, r_2$.
For the invariants we find
$$g_2={1\over 12}-{m^2\over 4L^2}r_s^2={1\over 12}-{r_s\over 4}{r_1+r_2+r_3\over
r_1r_2+r_1r_3+r_2r_3}\eqno(2.22)$$ 
$$g_3={1\over 6^3}-{m^2\over 48L^2}r_s^2-{a_0\over 16}r_s^2$$
$$={1\over 6^3}-{1\over 48}{(r_1+r_2+r_3)r_s\over r_1r_2+r_1r_3+r_2r_3}
+{1\over 16}{r_s^2\over r_1r_2+r_1r_3+r_2r_3}.\eqno(2.23)$$
Here $r_3$ has to be substituted by (2.21). For the derivatives $f'(r_1)$,
$f''(r_1)$ which appear in our solution (2.13) we obtain
$$f'(r_1)=-{r_1(r_1-r_2)(r_1-r_3)\over r_1r_2+r_1r_3+r_2r_3}\eqno(2.24)$$
$$f''(r_1)=-2{(r_1-r_2)r_1+(r_1-r_3)r_1+(r_1-r_2)(r_1-r_3)\over r_1r_2+r_1r_3+r_2r_3}.
\eqno(2.25)$$
With these substitutions the result (2.13) gives all possible geodesics in the form 
$r=r(\fii;r_1,r_2)$. This will be discussed in the next section.

As a first check of the solution (2.13) we consider the Newtonian limit. Let the two zeros
$r_1$, $r_2$ be real and large compared to the Schwarzschild radius $r_s$ in absolute value.
Then neglecting $O(r_s)$ in the invariants (2.22-23) the $\pe$-function becomes elementary
([2],p.652, equation 18.12.27):
$$\pe(\fii;{1\over 12},6^{-3})=-{1\over 12}+{1\over 4\sin^2\fii/2}=-{1\over 12}+
{1\over 2(1-\cos\fii)}.\eqno(2.26)$$
The leading order in the derivatives of $f$ is given by
$$f'(r_1)=-{r_1\over r_2}(r_1-r_2)$$
$$f''(r_1)=-2\Bigl(3{r_1\over r_2}-2\Bigl).\eqno(2.27)$$
It is convenient to introduce the eccentricity $\eps$ by
$${r_1\over r_2}={1-\eps\over 1+\eps}.\eqno(2.28)$$
Using all this in (2.13) we find the wellknown conic 
$$r={(1+\eps)r_1\over 1+\eps\cos\fii}.\eqno(2.29)$$
Assuming both zeros $r_1, r_2$ positive and $r_1<r_2$ we have $\eps<1$ and the orbit is
an ellipse with perihelion $r_1$ and aphelion $r_2$. In the hyperbolic case $\eps>1$ we
see from (2.28) that if $r_1$ is positive $r_2$ must be negative. Then there is only one
physical turning point $r_1$ which is the point of closest approach. The latter always
corresponds to $\fii=0$. The relativistic corrections to (2.29) are calculated in the
following section.

\section{Discussion of the solution}

The solution $r=r(\fii)$ (2.13) is an elliptic function of $\fii$ which implies that
it is doubly-periodic ([2], p.629 or any book on elliptic functions).
The values of the two half-periods $\omega, \omega'$ depend on the three roots of the fundamental 
cubic equation 
$$4e^3-g_2e-g_3=0.\eqno(3.1)$$
Again it is not necessary to solve this equation because the solutions $e_j, j=1,2,3$
can be easily obtained from the roots $0, r_1, r_2, r_3$ of our quartic $f(r)=0$. To see
this we transform $f(r)$ to Weierstrass' normal form as follows. First we set $r=1/x$
so that from (2.16) we get
$$f(r)={1\over x^4}(4a_3x^3+6a_2x^2+4a_1x+a_0)$$
Next we remove the quadratic term by introducing
$${1\over r}=x={1\over a_3}\Bigl(e-{a_2\over 2}\Bigl).\eqno(3.2)$$
This gives the normal form of Weierstrass
$$f(r)={a_3^2\over (e-\eh a_2)^4}(4e^3-g_2e-g_3),\eqno(3.3)$$
with the above invariants (2.14-15). That means roots of $f(r)$ are simply related to
roots of (3.1) by the transformation
$$e_j={a_3\over r_j}+{a_2\over 2}={r_s\over 4r_j}-{1\over 12}.\eqno(3.4)$$

The cubic equation (3.1) with real coefficients has either three real roots or one real
and two complex conjugated roots. The first case occurs if the discriminant
$$\triangle =g_2^3-27g_3^2\eqno(3.5)$$
is positive, in the second case $\triangle$ is negative. In terms of the roots $\triangle$
is given by ([2], p.629, equation 18.1.8)
$$\triangle=16(e_1-e_2)^2(e_2-e_3)^2(e_3-e_1)^2.\eqno(3.6)$$
The physically interesting orbits correspond to the first case of real roots.
If we have two complex conjugated zeros $r_2=r_1^*$ then (2.28) implies that the
eccentricity $\eps$ is imaginary. Such orbits have been discussed by Chandrasekhar
([4], p.111). Now we discuss the various cases.

\subsection{Bound orbits}

In this case we have two positive turning points $r_2>r_1>0$, consequently there are three
real roots $e_1>0>e_2>e_3$ given by
$$e_1={r_s\over 4r_3}-{1\over 12}={1\over 6}-{r_s\over 4}{r_1+r_2\over r_1r_2}
\eqno(3.7)$$
$$e_2=-{1\over 12}+{r_s\over 4r_1},\quad e_3=-{1\over 12}+{r_s\over 4r_2},
\eqno(3.8)$$
Our convention is chosen in agreement with [2]. The real half-period $\omega$ of the
$\pe$-function is given by ([2], p.549, equation 18.9.8)
$$\omega=\int\limits_{e_1}^\infty {dt\over\sqrt{4t^3-g_2t-g_3}}={K(k^2)\over \sqrt{e_1-e_3}}
\eqno(3.9)$$
where $K(k^2)$ is the complete elliptic integral of the first kind with parameter
$$k^2={e_2-e_3\over e_1-e_3}=r_s{r_2-r_1\over r_1r_2-r_s(2r_1+r_2)}$$
$$={r_s\over r_1}{2\eps\over 1+\eps}\Bigl(1-r_s{3-\eps\over 1+\eps}\Bigl)
^{-1}.\eqno(3.10)$$

As a first application let us give the post-Einsteinian correction to the orbital precession.
If $k^2$ (3.10) is small we can use the expansion ([2], p.591, equation 17.3.11)
$$K(k^2)={\pi\over 2}\Bigl[1+\Bigl({1\over 2}\Bigl)^2k^2+\Bigl({1\cdot 3\over 2\cdot 4}
\Bigl)^2k^4+\ldots\Bigl]\eqno(3.11)$$
From the roots $e_1, e_3$ we find
$${1\over\sqrt{e_1-e_3}}=2\Bigl[1+{r_s\over 2}{2r_1+r_2\over r_1r_2}+{3\over 8}r_s^2
\Bigl({2r_1+r_2\over r_1r_2}\Bigl)^2+O(r_s^3)\Bigl].$$
This finally leads to the half-period
$$\omega=\pi\Bigl\{1+{3r_s\over 4}{r_1+r_2\over r_1r_2}+{3r_s^2\over 8(r_1r_2)^2}
\Bigl[(2r_1+r_2)(2r_2+r_1)+{3\over 8}(r_2-r_1)^2\Bigl]\Bigl\}$$
$$=\pi\Bigl\{1+{3\over 2}{r_s\over r_1}{1\over 1+\eps}+{3\over 18}{r_s^2\over r_1^2}
{18+\eps^2\over (1+\eps)^2}+O(r_s^3)\Bigl\}\eqno(3.12)$$
The perihelion precession is given by $\triangle\fii=2(\omega-\pi)$. Then the order $r_s$
in (3.12) is Einstein's result and the $O(r_s^2)$ gives the correction to it. The accurate
computation of the half-period is necessary to control the orbit in the large. 

To compute the relativistic corrections for $r(\fii)$ from (2.13) we express the $\pe$-function
by Theta functions ([5], p.464)
$$\te_1(z,q)=2q^{1/4}(\sin z-q^2\sin 3z+q^6\sin 5z-\ldots)$$
$$\te_2(z,q)=2q^{1/4}(\cos z+q^2\cos 3z+q^6\cos 5z+\ldots)$$
$$\te_3(z,q)=1+2q(\cos 2z+q^3\cos 4z+q^8\cos 6z+\ldots)\eqno(3.13)$$
$$\te_4(z,q)=1-2q(\cos 2z-q^3\cos 4z+q^8\cos 6z-\ldots).$$
Here $q$ is the so-called Nome ([2], eq. 17.3.21)
$$q={k^2\over 16}+8\Bigl({k^2\over 16}\Bigl)^2+\ldots\eqno(3.14)$$
These series are rapidly converging since $k^2$ is small (3.10), {\it they give the
natural expansion in powers of the Schwarzschild radius $r_s$}. Now the $\pe$-function
is given in terms of Theta functions by ([2], eq. 18.10.5)
$$\pe(\fii)=e_2+{\pi^2\over 4\omega^2}\Bigl({\te'_1(0)\over\te_3(0)}{\te_3(\phi)\over
\te_1(\phi)}\Bigl)^2\eqno(3.15)$$
$$=-{1\over 12}+{r_s\over 4r_1}+{\pi^2\over 4\omega^2}{1\over\sin^2\phi}\Bigl[
1+4q(\cos^2\phi-1)+O(q^2)\Bigl],$$
where
$$\phi={\pi\over 2\omega}\fii.\eqno(3.16)$$
Using
$$f'(r_1)=2r_1\Bigl({\eps\over 1+\eps}-{r_s\over r_1}{3\eps+\eps^2\over (1+\eps)^2}\Bigl)$$
$$f''(r_1)=-2{1-5\eps\over 1+\eps}+6{r_s\over r_1}{1-4\eps-\eps^2\over (1+\eps)^2}
\eqno(3.17)$$
this leads to
$$\pe(\fii)-{f''(r_1)\over 24}={1+\eps-2\eps\sin^2\phi\over 4(1+\eps)\sin^2\phi}
\Bigl[1+{r_s\over r_1}{1\over 1+\eps\cos 2\phi}\Bigl(-3-{\eps\over 2}(1-\cos\phi)+$$
$$+2\eps{3+\eps\over 1+\eps}\sin^2\phi\Bigl)\Bigl]+O\Bigl({r_s\over r_1}\Bigl)^2.
\eqno(3.18)$$
Substituting this into (2.13) gives the desired orbit to $O(r_s)$
$${r(\fii)\over r_1}={1+\eps\over 1+\eps\cos 2\phi}+{r_s\over r_1}\eps{2\sin^2\phi\over
1+\eps\cos 2\phi}\Bigl\{{1\over 1+\eps\cos 2\phi}\Bigl[3+{\eps\over 2}(1-\cos\phi)-$$
$$-2\eps{3+\eps\over 1+\eps}\sin^2\phi\Bigl]-{3+\eps\over 1+\eps}\Bigl\}.\eqno(3.19)$$
It is important to insert the period $\omega$ in $\phi$ (3.16) according to (3.12) in
order to describe the perihelion precession correctly.

If the two roots $r_1=r_2$ coincide, it follows from (2.24) that $f'(r_1)=0$. According to
(2.13) we then have circular motion $r=r_1$. If all three zeros coincide $r_1=r_2=r_3$ then
(2.21)gives $r_3=3r_s$ which is the innermost circular orbit.

\subsection{Unbound orbits}

In this case there is only one physical point, the point of closest approach $r_1$.
The other root $r_2$ is negative, therefore, it is better to use the eccentricity $\eps$
(2.27) as the second basic quantity. With $r_3$ given by (2.21) we then have
$$r_1>r_3>0>r_2={1+\eps\over 1-\eps}r_1,\eqno(3.20)$$
because $\eps\ge 1$. The periodicity of (2.13) in $\fii$ is now realized by a jump to
an unphysical branch with $r<0$. In reality a comet moves on one branch only, but it is
a tricky problem to decide on which one. This is due to the fact that the period differs
a little from $2\pi$ as in the bounded case. Consequently, neighboring physical branches
$r>0$ are a little rotated against each other and the distinction between them is not easy. 
The quantity of physical interest is the direction $\fii_\infty$ of the 
asymptote. It follows from the original equation (2.11) by integrating the inverse over $r$
from $r_1$ to $\infty$
$$\fii_\infty=\int\limits_{r_1}^\infty {dr\over\sqrt{f(r)}}.\eqno(3.21)$$
This is an elliptic integral which can be transformed to Legendre's normal form
$$\fii_\infty={\mu\over\sqrt{a_0}} \int\limits_0^{\Phi_2}{d\Phi\over\sqrt{1-k^2\sin^2\phi}}
\eqno(3.22)$$
by the transformation ([6], vol.II, p.308)
$$\sin^2\Phi={r_3-r_2\over r_1-r_2}{r-r_1\over r-r_3},\quad
\sin^2\Phi_2={r_3-r_2\over r_1-r_2}.\eqno(3.23)$$
The parameter $k^2$ in (3.22) is given by
$$k^2={r_3\over r_1}{r_1-r_2\over r_3-r_2},\eqno(3.24)$$
and
$$\mu={2\over\sqrt{r_1(r_3-r_2)}}.\eqno(3.25)$$
The integral (3.22) is an incomplete elliptic integral of the first kind
$$\fii_\infty={\mu\over\sqrt{a_0}}F(\Phi_2,k^2)\eqno(3.26)$$
which has the expansion ([6], vol.II, p.313)
$$F(\Phi_2,k^2)=\Phi_2+{k^2\over 4}(\Phi_2-{1\over 2}\sin 2\Phi_2)+O(k^4).
\eqno(3.27)$$

For small $r_s/r_1$ we find
$$k^2={2\eps\over\eps+1}{r_s\over r_1}+O\Bigl({r_s\over r_1}\Bigl)^2$$
$${\mu\over\sqrt{a_0}}=2+{3-\eps\over 1+\eps}{r_s\over r_1}+O\Bigl({r_s\over r_1}\Bigl)^2$$
$$\sin^2\Phi_2={\eps+1\over 2\eps}\Bigl(1-{r_3\over r_1}\Bigl).$$
This gives
$$\cos 2\Phi_2=-{1\over\eps}-{\eps-1\over\eps}{r_3\over r_1}$$
and
$$\fii_\infty=2\Phi_2+{r_s\over r_1}\Bigl({3\over\eps+1}\Phi_2-{\eps\over 2\eps+2}
\sin 2\Phi_2\Bigl).\eqno(3.28)$$
It is convenient to calculate
$$\cos\fii_\infty=-{1\over\eps}-{r_s\over r_1}\Bigl({\eps-1\over 2\eps}+3{\sqrt{\eps^2-1}
\over\eps(\eps+1)}\Phi_2\Bigl)+O(r_s^2).\eqno(3.29)$$
The leading order is the Newtonian asymptote of the hyperbola.

\subsection{Null geodesics}

For $m^2=0$ there is only one constant of integration in the quartic (2.11)
$$f(r)={r^4\over d^2}-r^2+r_sr$$
which is the so-called impact parameter
$$d={L\over E}.\eqno(3.30)$$
Now it is necessary to calculate the roots of $f(r)=0$. This is easily done by means of a
power series expansion
$$r=c_0d+c_1r_s+c_2r_s^2+\ldots$$
We find
$$r_1=d\Bigl(1-{\delta\over 2}-{3\over 8}\delta^2+O(\delta^3)\Bigl)$$ 
$$r_2=-d\Bigl(1+{\delta\over 2}-{3\over 8}\delta^2+O(\delta^3)\Bigl)\eqno(3.31)$$ 
$$r_3=d(\delta+\delta^3),$$
where
$$\delta={r_s\over d}\eqno(3.32)$$
and we have ordered the zeros in the same way as in (3.20). Then as in the last subsection
we can calculate the direction of the asymptote (3.21) which now is equal to
$$\fii_\infty=\mu d F(\Phi_1,k^2)\eqno(3.33)$$
with  $\Phi_1$ given by (3.23)
$$\sin^2\Phi_1={1\over 2}+{3\over 4}\delta+O(\delta^3)\eqno(3.34)$$
and $k^2$ by (3.24)
$$k^2=2\delta\Bigl(1-\delta+{25\over 8}\delta^2\Bigl)\eqno(3.35)$$
and $\mu$ by (3.25)
$$\mu={2\over d}\Bigl(1-{\delta\over 2}+{9\over 8}\delta^2\Bigl).\eqno(3.36)$$

We want to calculate the light deflection in the post-Einsteinian approximation.
Using again the expansion (3.27) we have
$$\fii_\infty=2\Bigl(1-{\delta\over 2}+{9\over 8}\delta^2\Bigl)\Bigl[\Phi_1+{\delta\over 2}
(1-\delta)(\Phi_1-{1\over 2}\sin 2\Phi_1)\Bigl].$$
From (3.34) we obtain
$$\Phi_1={\pi\over 4}+{3\over 4}\delta+O(\delta^3).$$
Then up to $O(\delta^2)$ we find
$$\fii_\infty={\pi\over 2}+\delta+\Bigl({3\over 4}+{3\over 16}\pi\Bigl)\delta^2.
\eqno(3.37)$$
The deflection angle is given by
$$\triangle\fii=2\Bigl(\fii_\infty-{\pi\over 2}\Bigl)=2\delta+\Bigl({3\over 2}
+{3\over 8}\pi\Bigl)\delta^2.\eqno(3.38)$$
Instead of the impact parameter $d$ in $\delta$ (3.32) we would like to use the
distance of closest approach $r_1$ (2.24) in the form
$$\delta_1={r_s\over r_1}.\eqno(3.39)$$
The two are related by
$$\delta=\delta_1-{1\over 2}\delta_1^2+O(\delta_1^3)$$
which leads to
$$\triangle\fii=2\delta_1+\Bigl({1\over 2}
+{3\over 8}\pi\Bigl)\delta_1^2.\eqno(3.40)$$
The first term $2\delta_1$ is Einstein's result.

\section{The motion in time $t=t(\fii)$}

By dividing (2.7) by (2.10) we find
$${dt\over dr}=E{r\over r-r_s}\Bigl[E^2-\Bigl(1-{r_s\over r}\Bigl)\Bigl(m^2+{L^2\over r^2}
\Bigl)\Bigl]^{-1/2}$$
$$={E\over L}{r^3\over(r-r_s)\sqrt{f(r)}}.\eqno(4.1)$$
We choose $t=0$ at the point of closest approach $r=r_1$ and get
$$t={E\over L}\int\limits_{r_1}^r{dx\over\sqrt{f(x)}}\Bigl(x^2+r_sx+r_s^2+{r_s^3\over
x-r_s}\Bigl).\eqno(4.2)$$
This is a sum of elliptic integrals of first, second and third kind. Calculating these
gives the coordinate time as a function of $r$. However, we want $t$ as a function of
$\fii$ and, therefore, use the substitution (6.7) of the Appendix again
$$x=r_1+{f'(r_1)\over 4\pe(\fii)-f''(r_1)/6}\eqno(4.3)$$
$${dx\over\sqrt{f(x)}}=d\fii,$$
where the last relation follows from (2.11).

The last integral $O(r_s^3)$ in (4.2) is a small correction and we neglect it at the
moment. Then integrals of the following form remain to be calculated
$$J_n(\fii)=\int\limits_0^\fii{du\over (\pe(u)-\pe(v))^n}\eqno(4.4)$$
where we have set
$$\pe(v)={f''(r_1)\over 24}.\eqno(4.5)$$
Such integrals are known ([7], vol.4, p.109-110)
$$J_1(\fii)={1\over\pe'(v)}\Bigl[2\zeta(v)\fii+\log{\sigma(v-\fii)\over\sigma(v+\fii)}\Bigl]
\eqno(4.6)$$
$$J_2(\fii)=-{1\over\pe'^2(v)}\Bigl[\zeta(\fii+v)+\zeta(\fii-v)+2\pe(v)\fii+\pe''(v)J_1(\fii)
\Bigl].\eqno(4.7)$$
These results are easily verified by differentiating and using addition formulas.
Of course $J_0(\fii)$ is just the polar angle $\fii$. Then (4.2) leads to the
desired result for $t(\fii)$:
$$t(\fii)={E\over L}\Bigl\{\tau_0\fii+\tau_1J_1(\fii)
+\tau_2J_2(\fii)\Bigl\}+O(r_s^3).
\eqno(4.8)$$
where
$$\tau_0=r_1^2+r_sr_1+r_s^2\eqno(4.9)$$
$$\tau_1=\Bigl({r_1\over 2}+{r_s\over 4}\Bigl)f'(r_1)\eqno(4.10)$$
$$\tau_2={f'^2(r_1)\over 16}.\eqno(4.11)$$
Again we evaluate this for bounded orbits in the post-Newtonian approximation by means
of the expansion in Theta functions.

The quantity $v$ in (4.8-11) is given as the zero of (3.18). Introducing
$$V={\pi\over 2\omega}v\eqno(4.12)$$
we find
$$\cos 2V=-{1\over\eps}-{r_s\over r_1}{3\eps+1\over 2\eps}\equiv-\beta.\eqno(4.13)$$
Since $\eps<1$, $V$ is complex:
$$2V=\pi+i\log\Bigl(\beta+\sqrt{\beta^2-1}\Bigl)\equiv\pi+2ib\eqno(4.14)$$
Using ([2], eq.18.10.6)
$$\pe'(v)=-{\pi^3\over 4\omega^3}{\te_2(V)\te_3(V)\te_4(V)\te_1'^3(0)\over
\te_2(0)\te_3(0)\te_4(0)\te_1^3(V)}$$
$$=-{\pi^3\over 4\omega^3}{\cos V\over\sin^3 V}+O(q^2)$$
we obtain
$$\pe'(v)=i{\pi^3\over 2\omega^3}{(\beta^2-1)^{1/2}\over (\beta+1)^2}+O(r_s^2)$$
$$=i{\eps\over 2}{\sqrt{1-\eps^2}\over (1+\eps)^2}\Bigl(1+{r_s\over r_1}
{3\eps^2+4\eps-5\over 1-\eps^2}\Bigl).\eqno(4.15)$$
Similarly we calculate $\zeta(v)$ from ([2], eq.18.10.7)
$$\zeta(v)={\eta v\over\omega}+{\pi\over 2\omega}{\te_1'(V)\over\te_1(V)}\eqno(4.16)$$
where
$$\eta=\zeta(\omega)=-{\pi^2\over 12\omega}{\te_1'''(0)\over\te_1'(0)}\eqno(4.17)$$
and $\sigma(z)$ from ([2], eq.18.10.8)
$$\sigma(z)={2\omega\over\pi}\exp\Bigl({\eta z^2\over 2\omega}\Bigl){\te_1(Z)\over\te_1'(0)},
\quad Z={\pi z\over 2\omega}.\eqno(4.18)$$
This implies
$$\log{\sigma(v-\fii)\over\sigma(v+\fii)}=-2{\eta\over\omega}v\fii+\log{\sin(V-\Phi)\over\sin(V+\Phi)}
+O(q^2)$$
$$=-2{\eta\over\omega}v\fii+\log{\cos (ib-\Phi)\over\cos (ib+\Phi)}\equiv -2{\eta\over\omega}v\fii+2i\alpha(\fii),\eqno(4.19)$$
where $\alpha$ is given by
$$\alpha(\fii)=\arctan (\tan\Phi\tanh b).\eqno(4.20)$$
Then $J_1(\fii)$ (4.6) is equal to
$$J_1(\fii)=-{2\omega^2\over\pi^2}(\beta+1)\fii+{4\omega^3\over\pi^3}{(\beta+1)^2
\over\sqrt{\beta^2-1}}\alpha(\fii)+O\Bigl({r_s\over r_1}\Bigl)^2.$$

To expand this in the post-Newtonian order we first calculate $\alpha(\fii)$
from
$$\tan\alpha=\sqrt{{1-\eps\over 1+\eps}}\tan\Phi+{r_s\over r_1}\eps{1+3\eps\over
1-\eps^2}\sqrt{{1-\eps\over 1+\eps}}\tan\Phi.$$
Introducing
$$\psi=2\arctan\Bigl(\sqrt{{1-\eps\over 1+\eps}}\tan \Phi\Bigl)\eqno(4.21)$$
we get
$$\alpha={\psi\over 2}+{r_s\over r_1}{\eps\over 4}{1+3\eps\over 1-\eps^2}\sin\psi.$$
As before in (3.19) we do not expand $\Phi$ in (4.21). However, if one does so one
finds a contribution $O(r_s/r_1)$
$$\psi=2\arctan\Bigl(\sqrt{{1-\eps\over 1+\eps}}\tan (\fii/2)\Bigl)-
{3\over 2}{r_s\over r_1}\sqrt{{1-\eps\over 1+\eps}}{\fii\over 1+\eps\cos\fii}
\eqno(4.22)$$
where the first term, say $\psi_N$, is the parameter which appears in Newtonian
mechanics (Kepler's equation, see below (4.29)). Now the expansion of $J_1(\fii)$ is
given by
$$J_1(\fii)=J_1^0(\fii)+{r_s\over r_1}J_1^1(\fii)$$
$$J_1^0(\fii)=2{1+\eps\over\eps}\Bigl(\psi\sqrt{{1+\eps\over 1-\eps}}-\fii\Bigl)\eqno(4.24)$$
$$J_1^1(\fii)={1\over\eps}\Bigl[-\fii(7+3\eps)+\eps){1+3\eps\over 1-\eps^2}
\sqrt{{1+\eps\over 1-\eps}}\sin\psi+$$
$$+2\psi\sqrt{1+\eps\over 1-\eps}{5-4\eps-3\eps^2\over 1-\eps}\Bigl].$$
For $\eps\to 0$ we have the simple finite limit
$$J_1(\fii)\vert_{\eps=0}={2\omega\over\pi}\Bigl(1+3{r_s\over r_1}\Bigl)
(2\Phi-\sin 2\Phi)$$
which also follows directly from the definition (4.4).

To expand
$$J_2(\fii)=-{1\over\pe'^2(v)}\Bigl[2({\eta\over\omega}+\pe(v))\fii-{\pi\over\omega}
{\sin 2\Phi\over\cos 2\Phi+\beta}+\pe''(v)J_1(\fii)\Bigl]$$
$$=J_2^0(\fii)+{r_s\over r_1}J_2^1(\fii)$$
we need
$$\pe''(v)=6\pe^2(v)-{g_2\over 2}$$
$$={\eps\over 2}{2\eps-1\over (1+\eps)^2}\Bigl(1+{r_s\over 2r_1}{11-18\eps-5\eps^2\over
(2\eps-1)(1+\eps)}\Bigl).$$
This finally leads to
$$J_2^0(\fii)={4\over\eps}{(1+\eps)^2\over 1-\eps}\Bigl[\fii-{(1+\eps)\sin 2\Phi
\over 1+\eps\cos 2\Phi}+{2\eps)-1\over 2(1+\eps)}J_1^0(\fii)\Bigl]\eqno(4.25)$$
$$J_2^1(\fii)={2\over\eps}{(1+\eps)^2\over 1-\eps}\Bigl[\fii{15-12\eps-11\eps^2
\over 1-\eps^2}-$$
$$-{\sin 2\Phi\over 1+\eps\cos 2\Phi}\Bigl({1+4\eps+3\eps^2\over 1+\eps\cos 2\Phi}
+{17-13\eps-12\eps^2\over 1-\eps}\Bigl)+$$
$$+J_1^0(\fii){6-15\eps+8\eps^2-\eps^3\over (1+\eps)^2(1-\eps)}
-J_1^1(\fii){1-2\eps\over 1+\eps}\Bigl].\eqno(4.26)$$
Again we do not expand $\Phi$ (3.16) in order to keep the perihelion precession
as precise as possible. The limit for $\eps\to 0$ is equal to
$$J_2(\fii)\vert_{\eps=0}=2{\omega\over\pi}\Bigl(1+6{r_s\over r_1}\Bigl)
(6\Phi-4\sin 2\Phi+\sin 2\Phi \cos 2\Phi).$$

In the final result (4.8) for the time
$$t(\fii)=t_0(\fii)+{r_s\over r_1}t_1(\fii)\eqno(4.27)$$
the pre-factor $E/L$ also gives a correction:
$${E\over L}=\sqrt{{2\over r_sr_1(1+\eps)}}\Bigl(1-{r_s\over r_1(1+\eps)}\Bigl)$$
which follows from (2.19-21). In $t_0$ the terms proportional to $\fii$ cancel
$$t_0(\fii)=\sqrt{{2r_1^3\over r_s(1+\eps)}}{1+\eps\over 1-\eps}\Bigl({\psi\over
\sqrt{1-\eps^2}}-{\eps\sin 2\Phi\over 1+\eps\cos 2\Phi}\Bigl).\eqno(4.28)$$
Approximating $2\Phi$ by $\fii$ this is in agreement with Kepler's equation
$$t_0(\fii)=\sqrt{{2r_1^3\over r_s(1-\eps)^3}}(\psi
-\eps\sin \psi).\eqno(4.29)$$
The post-Newtonian corrections in (4.27) come from various places. To show this we
write the result in the form
$$t_1(\fii)=t_0(\fii){\eps\over 1+\eps}+\eps\sqrt{{2r_1^3\over r_s(1+\eps)}}\Bigl[
{1\over 1+\eps}J_1^1(\fii)-$$
$$-{1\over 2}{\eps+5\over (1+\eps)^2}J_1^0(\fii)+{\eps\over 4(1+\eps^2)}J_2^1(\fii)-
{\eps\over 2}{3+\eps\over (1+\eps)^3}J_2^0(\fii)\Bigl].\eqno(4.30)$$
As in (3.19) the post-Newtonian correction vanishes for circular motion $\eps=0$.

\section{Gravitational red shift}

The study of Schwarzschild geodesics is relevant for the investigation of the recently
discovered S-stars near the Galactic Center ([1] and references given there). These stars
move in the strong gravitational field of the central black hole so that general relativistic
effects are observable and the Schwarzschild metric $g_{\mu\nu}$ is a fairly good
description of the situation. The measurable quantity of interest is the red shift of
spectral lines in the light emitted by the moving star. Therefore we finally consider this.

Let $\nu_1$ be the frequency of a given atomic line from the star and $\nu_0$ the frequency
of the same line observed in the rest frame of the galaxy. If $dx^\mu/dt$ is the velocity
of the star, the two frequencies are related by ([9] p.83, equ. 3.5.6)
$${\nu_1\over\nu_0}={\Bigl(g_{\al\beta}(x){dx^\al\over dt}{dx^\beta\over dt}\Bigl)^{1/2}
\over g_{00}(X)^{1/2}}.\eqno(5.1)$$
We assume that the observer at $X$ is far away from the center such that the denominator
can be approximated by 1. For a star moving in the plane $\te=\pi/2$ we have
$$g_{\al\beta}(x){dx^\al\over dt}{dx^\beta\over dt}=e^\nu-e^\nu\Bigl({dr\over dt}\Bigl)^2
-r^2\Bigl({d\fii\over dt}\Bigl)^2.\eqno(5.2)$$
From (2.10) and (2.7) we find
$${dr\over dt}=e^\nu\sqrt{1-{e^\nu\over E^2}\Bigl(m^2+{L^2\over r^2}\Bigl)}$$
and (2.8) gives
$${d\fii\over dt}={L\over E}{e^\nu\over r^2}.$$
Substituting all this into (5.1) we see that $L$ drops out and we end up with the simple
result
$${\nu_1\over\nu_0}={m\over E}e^\nu={m\over E}\Bigl(1-{r_s\over r(\fii)}\Bigl).
\eqno(5.3)$$

By (2.20) we can express $E$ by the perihelion $r_1$ and aphelion $r_2$
$$E^2=m^2\Bigl(1-{r_s\over r_1+r_2+r_3}\Bigl)$$
where $r_3$ is the small correction (2.21). Then we finally get
$${\nu_1\over\nu_0}=\Bigl(1-{r_s\over r(\fii)}\Bigl)\Bigl(1-{r_s\over r_1+r_2+r_3}
\Bigl)^{-1/2}.\eqno(5.4)$$
The lowest order $O(r_s)$ is equal to
$${\nu_1\over\nu_0}=1-{r_s\over r(\fii)}+{r_s\over 2(r_1+r_2)}+O(r_s^2).\eqno(5.5)$$
Since the last term is always smaller than the second one we indeed have red shift $\lambda_1
>\lambda_0$. Of course, it is maximal at the perihelion where $r=r_1$ is minimal. The
total observed red shift is obtained by multiplying (5.5) with the Doppler factor
$(1+v_r)^{-1}$ where $v_r$ is the component of the relative velocity along the
direction from the observer to the star ([9], p.30).

{\bf Acknowledgment}. It is a pleasure to acknowledge elucidating discussions with
Prasenjit Saha, in particular the introduction into the fascinating field of Galactic-center
stars. I also thank Raymond Ang\'elil for showing his simulations of the corresponding
dynamics.

\section{Appendix: Integration of the differential equation}

We closely follow Whittaker and Watson ([5], p.452). With the notation of the paper (2.11), let
$$\fii=\int\limits_{r_1}^r {dx\over\sqrt{f(x)}}\eqno(6.1)$$
where $r_1$ is any zero, $f(r_1)=0$. By Taylor's theorem, we have
$$f(x)=4A_3(x-r_1)+6A_2(x-r_1)^2+4A_1(x-r_1)^3+A_0(x-r_1)^4,$$
where
$$A_0=a_0,\quad A_1=a_0r_1+a_1$$
$$A_2=a_or_1^2+2a_1r_1+a_2,\quad A_3=a_0r_1^3+3a_1r_1^2+3a_2r_1+a_3.$$
Introducing the new integration variable
$$s=(x-r_1)^{-1},\quad s_1=(r-r_1)^{-1},\eqno(6.2)$$
we have
$$\fii=\int\limits_{s_1}^\infty [4A_3s^3+6A_2s^2+4A_1s+A_0]^{-1/2}ds.$$
To remove the second term in the cubic we set
$$s={1\over A_3}(z-{A_2\over 2}),\quad s_1={1\over A_3}(z_1-{A_2\over 2})\eqno(6.3)$$
and we get
$$\fii=\int\limits_{z_1}^\infty [4z^3-(3A_2^2-4A_1A_3)z-(2A_1A_2A_3-A_2^3-A_0A_3^2)]^{-1/2}dx
\eqno(6.4)$$
The coefficients of $z$ and $z^0$ are just the invariants $g_2, g_3$ (2.14-15) of the
original quartic.

Now the inversion of the integral gives Weierstrass' $\pe$-function
$$z_1=\pe (\fii;g_2,g_3).\eqno(6.5)$$
From (6.2) and (6.3) we have
$$r=r_1+{A_3\over z_1-A_2/2}\eqno(6.6)$$
and hence
$$r=r_1+{f'(r_1)\over 4\pe(\fii)-f''(r_1)/6}.\eqno(6.7)$$

\end{document}